\documentclass[aps, prb,superscriptaddress,floatfix]{revtex4}
\usepackage{graphicx}
\usepackage{dcolumn}
\usepackage{bm}
\usepackage{natbib}
\usepackage{upgreek}
\usepackage{amsmath}
\usepackage{amssymb}

\newcommand*{\Dy}{DyMnO$_3$}

\begin{document}

\title{Order-disorder type critical behaviour at the magnetoelectric phase transition in multiferroic \Dy}

\author{M. Schiebl}
\author{A. Shuvaev}
\author{Anna Pimenov}
\author{G. E.  Johnstone}
\author{V. Dziom}
\affiliation{Institute of Solid State Physics, Vienna University of
Technology, A-1040 Vienna, Austria}
\author{A. A. Mukhin}
\author{V.Yu. Ivanov}
\affiliation{Prokhorov General Physics Institute, Russian Academy of
Sciences, 119991 Moscow, Russia}
\author{A. Pimenov}
\affiliation{Institute of Solid State Physics, Vienna University of
Technology, A-1040 Vienna, Austria}

\begin{abstract}

We present the results of detailed dielectric investigations of the  relaxation dynamics
in \Dy~ multiferroic manganite. Strong low-frequency relaxation process near the paraelectric-ferroelectric
phase transition is observed.
We provide an experimental evidence that this
relaxation mode corresponds to a chirality switching of the spin
cycloid in \Dy. We demonstrate that the relaxation dynamics in \Dy~ is
typical for an order-disorder phase transition and may be understood within a simple model with a double well potential.
These results suggest the interpretation of the
paraelectric sinusoidal phase in manganites as a dynamical equilibrium of magnetic cycloids with
opposite chiralities.

\end{abstract}

\date{\today}


\maketitle

\section{Introduction}

Multiferroic materials with a coupling of electric and magnetic degrees
of freedom have attracted considerable interest after the discovery
of a large magnetoelectric effect (ME) in several
compounds~\cite{fiebig_jpd_2005,ramesh_nmat_2007,%
tokura_science_2006,cheong_nmat_2007}. They are currently the subject of
intensive study due to fascinating physical properties and potential for
applications as multifunctional devices~\cite{Spaldin15072005,eerenstein_nature_2006}.
The rare-earth RMnO$_{3}$ manganites (R=Gd, Tb, Dy, Eu/Y) with
orthorhombically distorted perovskite structure have emerged as a new class
of multiferroics with strongly coupled antiferromagnetic and ferroelectric
properties~\cite{kimura_prb_2005,hemberger_prb_2007}.
Several rare earth manganites orders antiferromagnetically below $T_N \sim 40$~K into collinear paraelectric phase\cite{kenzelmann_prl_2005} with sinusoidal modulations of spins. This phase is followed by a cycloidal spin order with nonzero electric polarization below $T_C \sim 20$~K.

It should be
noted that the Mn$^{3+}$ atom is supposed~\cite{Mochizuki_2009PRB} to have
a Heisenberg spin with a fixed length of $S = 2$. The purely
sinusoidally modulated spin phase contradicts this property. Model calculations~\cite{Mochizuki_2009PRB} have
obtained the sinusoidal order only as a time-space average of the
simulated cluster. Thus, there is a possibility that a short-range
dynamic order exists in the intermediate temperature range $T_{C} <
T < T_{N}$ which is responsible for the ``hidden'' spin. As the
magnetic order at low temperatures is spin cycloid, it is natural to
assume that the dynamic short-range order is also a spin cycloid.
This would imply that there are fluctuating ferroelectric regions in
the sinusoidal phase and that the ferroelectric transition is
actually of the order-disorder type. Such transition has been also suggested in Ref.~[\onlinecite{schrettle_prl_2009}] where a c-axis relaxation typical
for the order-disorder type transitions has been investigated. The fact that the wave
vector $q_{Mn}$ of the spin wave does not change at the transition
temperature $T_C$~[\onlinecite{Arima_2006PRL}] is also an indirect
evidence that the ferroelectric transition is not of the displacive
type.

The data by terahertz spectroscopy~\cite{pimenov_jpcm_2008,pimenov_prb_2008}
evidence the nonzero dielectric contribution of electromagnon in the
sinusoidal phase. According to the commonly accepted mechanism of
the electromagnon~\cite{aguilar_prl_2009,lee_prb_2009}, the majority of the spectral weight of this mode originates from exchange striction mechanism and can
only exist in magnetic phases with non-collinear spin arrangement.
These facts again favor the hypothesis of dynamical cycloidal spin order in the
sinusoidal phase.

Recent theoretical analysis of the terahertz dynamics in the sinusoidal phase suggested an explanation based on anomalous magnetoelectric coupling. Investigations of the collinear sinusoidal phase
in the diluted compounds TbMn$_{1-x}$Al$_x$O$_3$~[\onlinecite{Cuartero_PRB2012}] and the observation
of the memory effect in the low temperature sinusoidal phase in the
multiferroic MnWO$_4$~[\onlinecite{Taniguchi_PRL2009}] have
suggested the presence of the nanosize ferroelectric domains and
support relaxor order-disorder type transition.

In this work we present the analysis of the critical behavior of the low-frequency
relaxation in \Dy{}. The observed critical behavior confirms
that the sinusoidal to cycloidal phase transition is of the
order-disorder type. Our model suggests the presence of the short-range cycloidal order in the collinear spin phase.

\subsection{\Dy}

In zero magnetic field, \Dy{} undergoes an antiferromagnetic (AFM)
transition with a temperature dependent modulation vector (0
$q_{Mn}$ 1) around $T_{N} =39$~K. With further cooling, the $q_{Mn}$ value is
locked at the transition temperature, $T_{C} \approx 19$~K, and
ferroelectric (FE) polarization appears simultaneously along the
$c$-axis. It is well established that the ferroelectric polarization
in \Dy{} is induced by a cycloidal magnetic
order~\cite{kenzelmann_prl_2005,Arima_2006PRL,Mochizuki_2009PRB}
through
the inverse Dzyaloshinskii-Moriya (DM) interaction~\cite{katsura_prl_2005,%
mostovoy_prl_2006,sergienko_prb_2006,cheong_nmat_2007} and it can be written as
\begin{equation}\label{EquPol}
  P=\sum_{i,j}Ke_{i,j}\times(S_{i}\times S_{j}) \ ,
\end{equation}
where $e_{i,j}$ denotes the unit vector connecting the spins $S_{i}$ and $S_{j}$, $K$ is a constant representing the exchange interaction and the spin-orbit interaction.
Accordingly, the electric polarization, $P$, is intimately linked to the chirality of the
magnetic cycloid (clockwise respectively counterclockwise cycloidal magnetic
ordering), that is, changing the direction of $+P\rightarrow-P$ implies
changing the rotation (chirality) of the magnetic cycloid. This was
demonstrated\cite{Fabrizi_2009PRL} by the asymmetry
in the scattering of left-hand and right-hand circularly polarized x-rays
by nonresonant magnetic x-ray diffraction for closely similar compound TbMnO$_{3}$.

The unique cycloidal magnetic ordering below $T_C$ in
RMnO$_{3}$ perovsikes is assigned to the Mn-3d
spins~\cite{kenzelmann_prl_2005,Voigt_2007PRB,Mannix_2007PRB,%
kimura_nature_2003,goto_prl_2004,kimura_prb_2003,kimura_prb_2005} but
ordering of the Dy-4f moments is also of interest~\cite{kenzelmann_prl_2005,%
Prokhnenko_2007PRL,Prokhnenko_2007PRL_TbMnO3}. Although it is the magnetic
structure of Mn subsystem that determines the emergence of ferroelectricity
in rare-earth manganites~\cite{Prokhnenko_2007PRL_TbMnO3}, the mutual
coupling of the Mn-3d and Dy-4f moments and, consequently, the ordering of
the 4f moments in Dy{} causes a particular large polarization observed
in this material~\cite{kimura_prb_2005,Aliouane_2008JPCM,Schierle_2010PRL}.
In addition, the basal plane of the spin cycloid flops from the $bc$ plane
to the $ab$ plane and rotates the polarization by applying a magnetic
field along the $b$-axis.

Concerning the temperature range $T_{C} < T < T_{N}$, there have
been a number of studies of the magnetic structure of \Dy{} and the
related material TbMnO$_{3}$ using techniques such as
neutron diffraction~\cite{quezel_physica_1977,kenzelmann_prl_2005}
and magnetic x-ray scattering~\cite{Mannix_2007PRB,Wilkins_PRL2009}.
These studies have all shown that there is a long-range
sinusoidal magnetic ordering of the Mn$^{3+}$ ions in this
temperature range, with the moments aligned parallel to the
crystallographic $b-$axis, although
Mannix~et~al.~[\onlinecite{Mannix_2007PRB}] and
Wilkins~et~al.~[\onlinecite{Wilkins_PRL2009}], both show a small
component of the moment is aligned with the $c-$axis.


\section{Experiment}

\Dy~ single crystals were grown in Ar flow by a
floating-zone method with radiation heating~\cite{balbashov_jcg_1996, mori_ml_2000}.
Terahertz properties of the samples from the same batch have been presented previously~\cite{shuvaev_prb_2010, shuvaev_prl_2013}.
The complex dielectric constant was measured for electric field along the crystallographic $a$-axis in the
frequency range 0.1Hz-1MHz using a frequency response analyzer in
magnetic fields 0-14 T and with  $B \| b$-axis.
We investigated the temperatures
near the critical temperature of the paraelectric-ferroelectric phase transition (red circles in Fig.~\ref{PD}) with an
increment of $\Delta T=0.1$~K by using a Physical Property Measurement
System (PPMS). Silver paint contacts were applied to the sample forming a capacitor.


\section{Pseudo-Spin-Model}

The theory of dynamic critical phenomena for a multivariable  system can be formulated as a
generalization of the single particle Langevin equation. The  equation of motion for the time-dependent local
configuration of the order parameter field is most conveniently given by the
time-dependent Ginzburg-Landau equation\cite{Nishimori_2011,Kardar_2007,Hohenberg_1977}.
\begin{equation}\label{EquPSM1}
  \frac{\partial P(\textbf{r},t)}{\partial t}=-\Gamma\frac{\delta F}{\delta P(\textbf{r},t)}+\zeta(\textbf{r},t) \ .
\end{equation}
Here, the  order parameter is given by static electric polarization $P$, $\Gamma$ is a dissipation
parameter, and $\zeta(\textbf{r},t)$ is a random noise simulating the effect of thermal
excitation of the order parameter. In order to guarantee that the system reaches the
canonical equilibrium probability distribution at long times, $\zeta(\textbf{r},t)$ is a
random Gaussian variable satisfying $\langle\zeta(\textbf{r},t)\rangle=0$ and
$\langle\zeta(\textbf{\'{r}},t)\zeta(\textbf{r},\acute{t})\rangle=2Dk_{B}T\delta(\textbf{r}-\textbf{\'{r}})\delta(t-\acute{t})$
[\onlinecite{Tauber_2014}], where $D$ is the diffusion coefficient. The crucial point is
to find an expression for the free energy in Eq.~(\ref{EquPSM1}). In the framework
of the Landau theory for a second order phase transition, Smolenskii
\cite{smolenskii_ufn_1982} introduced a
bi-quadratic term ($F\propto\gamma P^{2}M^{2}$, magnetodielectric effect) into the free energy which accounts for the
coupling between magnetization and electric polarization. Bi-quadratic terms are invariant to all symmetry operations
and thus they are allowed in any material with coupled spin and charge
degrees of freedom.  Since the dielectric susceptibility is determined by taking the second
derivative of the free energy with respect to the polarization, the dielectric constant
will be proportional to the square of the order parameter, $\varepsilon\propto M^{2}$
[\onlinecite{Kimura_2003}]. Describing magnetodielectric effects in antiferromagnetic
materials, the expression $F\propto\gamma P^{2}M^{2}$ is not sufficient since the
magnetization, $M$, remains zero in the ordered phase. In such a case, $M$ is replaced by
the antiferromagnetic vector, $L=M_1-M_2$. Here $M_1$ and $M_2$ are the magnetizations  of two antiferromagnetic subsystems.

Within a more general model, Lawes \cite{Lawes_2003PRL,Lawes_2009SSC}
et al. proposed the coupling of the polarization to the $q$-dependent magnetic
correlation function $\langle M_{q}M_{-q}\rangle$. This coupling leads to a magnetodielectric
term in the free energy  $F\propto\sum_{q}g(q)P^{2}\langle
M_{q}M_{-q}\rangle$, where $g(q)$ is a $q$-dependent coupling constant. The
$q$-dependence of the free energy via a spin-spin correlation function enables to apply
it to very general forms of magnetic order, including ferromagnetic (FM) and antiferromagnetic (AFM) transitions.
In order to obtain a microscopic theory for $g(q)$
in systems with a strong spin-lattice interaction, the coupling between the polarization and the spin correlations arises from the coupling of magnetic fluctuations to the optical phonons. That is, the spin correlations perturb the optical phonon frequencies which
in turn shift the dielectric constant through the spectral weight transfer and the Lyddane-Sachs-Teller relation. The model determines the coupling $g(q)$ by expanding the exchange integral of neighboring spins in terms of the normal coordinates for the phonons. Physically, this procedure corresponds to a coupling between the magnetic correlation function and atomic displacements.

In multiferroic rare earth manganites, RMnO$_3$, the electric polarization is directly linked to the chirality of the magnetic cycloid \cite{cheong_nmat_2007,mostovoy_prl_2006}.
Based on this fact, we propose that the polarization in the
ferroelectric phase in \Dy{} is proportional to the difference of opposite chiralities of
Mn$^{3+}$ magnetic cycloids. Here we assume an order-disorder type phase transition
between paraelectric and ferroelectric states. Similar analysis in a triangular lattice antiferromagnet RbFe(MoO$_{4})$ demonstrated a proportionality between polarization in the multiferroic phase and the chirality difference of the magnetic structure \cite{Kenzelmann_2007PRL}.

\begin{figure}[h]
\begin{center}
\includegraphics[width=0.5\linewidth, clip]{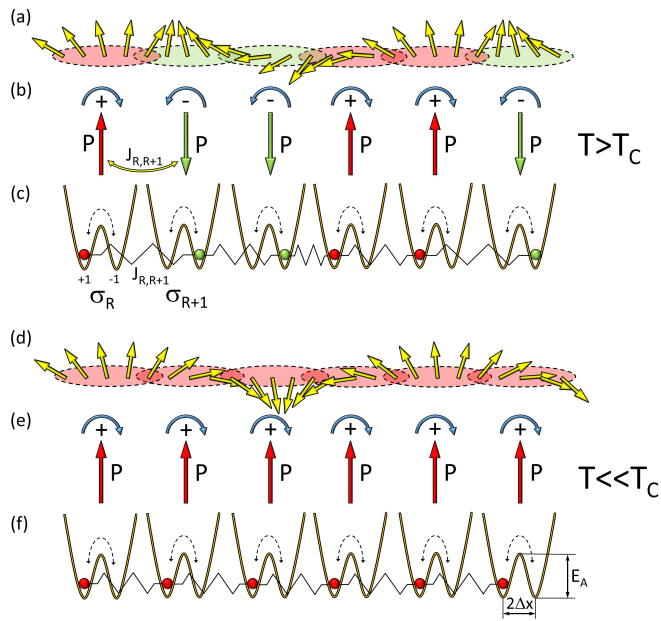}
\end{center}
\caption{\emph{Pseudo-Spin Model} (a) Short range cycloidal order of the Mn$^{3+}$ magnetic moments for $T > T_{C}$. (b) According to Eq.~(\ref{EquPol}), electric dipoles associated with the O$^{2-}$ ions are generated by the canting of neighboring spins leading to a mesoscopic electric polarization (red and green zones), however, the macroscopic electric polarization is zero for $T > T_{C}$. Each electric dipole interacts with neighboring electric dipoles. (c) Ising type pseudo spins in local double-well potential separating energetically the clockwise $ab$-cycloidal and the counterclockwise $ab$-cycloidal order of the Mn$^{3+}$ magnetic moments. Each pseudo spin is in a local double-well potential and interacts with neighboring pseudo spins by harmonic forces, represented as springs. Each pseudo spin generates a electric dipole moment of $\mu_{R}=\sigma_{R}q\Delta x$. Although the microscopic mechanism of the magnetoelectric coupling is more complex, within present simple model it is represented as an interaction between neighboring pseudo spins via harmonic forces.
(d) Long range cycloidal order of the Mn$^{3+}$ magnetic moments for $T\ll T_{C}$ and leading to (e) non-zero macroscopic polarization. (f) At low temperatures most of the pseudo spins occupy the same side of the double-well potential.}
\label{Phi4Model}
\end{figure}

In the present model the following assumptions for the phase transition in
\Dy~ are imposed: (i) A disorder between clockwise and counterclockwise $ab$-cycloidal
order of the Mn$^{3+}$ magnetic moments is assumed, (ii) The electric dipole moments are associated with the displacement of the O$^{2-}$ ions due to inverse DM interaction~\cite{cheong_nmat_2007,mostovoy_prl_2006}, (iii) The direction of the
electric dipoles depend on the chirality of the magnetic order, (iv) Two possible
direction of the electric dipoles are energetically separated by an energy barrier, (v)
Similar to Lawes\cite{Lawes_2003PRL,Lawes_2009SSC} et al. we propose a coupling of the magnetic correlation function and the correlation of O$^{2-}$ atomic displacements. Thus we assume that we can describe the
ordering of the magnetic sublattice by the ordering process of the O$^{2-}$ ions.

A useful tool for studying phase transitions is provided by the unidimensional $\phi^{4}$ single ion model
\cite{Bruce_1981,Giddy_1989JPCM,Salje_1991ACSA,Padlewski_1992JPCM,Radescu_1995JPCM}. This
model contains an array of atoms linked by harmonic forces, with one atom in each unit
cell (Fig.\ref{Phi4Model}). Each atom is located in its local double-well potential
which represents the rest of the crystal.

Assuming order-disorder limit ($E_{A}/k_{B}T_{C}\gg1$ [\onlinecite{Dove_1997,Aubry_JCP1975}]) the system can be described by the
pseudo spin formalism and thus the model Hamiltonian is essentially governed by an Ising
type interaction in combination with an interaction of dipoles with a homogeneous
electric field \cite{Dove_1997,Aubry_JCP1975,Strukov_1998}
\begin{equation}\label{EquPSM2}
  H \propto-\sum_{R,\acute{R}}\Delta x^{2}J_{R,\acute{R}}\sigma_{R}\sigma_{\acute{R}}-\sum_{R}E(t)q\Delta x\sigma_{R} \ ,
\end{equation}
where $\Delta x$ is the displacement of the O$^{2-}$ ion, $J_{R,\acute{R}}$ is the
coupling constant between O$^{2-}$ ions at position $R$ and $\acute{R}$,  $E(t)$ is a
time dependent homogeneous electric field, $q$ is the charge of the oxygen ion, and
$\sigma_{R}$ is the pseudo spin at position $R$ with $\sigma_{R}=x_{R}/\Delta x$ which can take
the values +1 and -1.

Introducing a statistical mean value of the spin variable,
$s=(N_{+}-N_{-})/(N_{+}+N_{-})$, with $N_{+}$ and $N_{-}$ being the occupation numbers of
O$^{2-}$ ions in +1 und -1 state respectively, and neglecting fluctuations of
correlations, the free energy of the system with polarization $P=nq\Delta x s$ is given by\cite{Strukov_1998}
\begin{widetext}
\begin{equation}\label{EquPSM3}
  F=-k_{B}T\ln Z=
  \int \left\{\frac{J_{0}}{nq^{2}\Delta x}P^{2}-
  nk_{B}T\ln\left[2\cosh\left(\frac{\frac{2\Delta x J_{0}}{nq}P+E(t)q\Delta x}{k_{B}T}\right)\right]\right\}dV \ ,
\end{equation}
\end{widetext}
where $n$ is the dipole density  and $J_{0}=\sum_{\acute{R}}J_{R,\acute{R}}$
represents the coupling constant characterizing the interaction of an
O$^{2-}$ ion at position $R$ with another O$^{2-}$ ion at position $\acute{R}$
located within an interaction radius. Since the first term in Eq. (\ref{EquPSM2}) is bilinear in the displacements of O$^{2-}$ ions it corresponds to a electric dipole-dipole interaction.

Expanding $F$ with respect to $P$ and $E$ and
taking the functional derivative of $F$ with respect to $P$, leads to an equation of
motion of the homogeneous order parameter:
\begin{widetext}
\begin{equation}\label{EquPSM4}
  \frac{\partial  P(t)}{\partial t}=-\Gamma\frac{\delta F}{\delta P}=
  -\Gamma\left[2\left(\frac{J_{0}}{nq^{2}}-\frac{2(\Delta x)^{2}J_{0}^{2}}{nq^{2}k_{B}T}\right)P+
  \frac{32(\Delta x)^{4}J_{0}^{4}}{6n^{3}q^{4}k_{B}^{3}T^{3}}P^{3}-\frac{2(\Delta x)^{2}J_{0}}{k_{B}T}E(t)\right] \ .
\end{equation}
\end{widetext}
Here $\Gamma=nq^{2}J_{0}^{-1}\nu_{0}\exp(-E_{A}/k_{B}T)$ is the damping parameter.

According to  Landau theory\cite{Landau_Book5} the phase transitions takes place when
the term in Eq.~(\ref{EquPSM4}) linear in $P$ vanishes. This determines the phase
transition temperature as $T_{C}=2J_{0}(\Delta x)^{2}/k_{B}$. Substituting $E(t)=\delta
E\exp(-i\omega t)$ and $P(t)=P_{S}+\delta P\exp(-i\omega t)$ [\onlinecite{Lines}], where
$P_{S}$ is the static electric polarization, and comparing the relaxation rate in a
single double well potential\cite{Strukov_1998,kremer_2002,Gonzalo} we get an expression for the relaxation strength $\Delta\varepsilon_{r}=\delta P/(\varepsilon_{0}\delta E)$ and the relaxation time $\tau$ in close agreement with Lines and Glass [\onlinecite{Lines}] and Blinc and \v{Z}ek\v{s} [\onlinecite{blinc_book}] as
\begin{equation}\label{EquPSM5}
  \Delta\varepsilon_{r}^{-1}=\left\{ \begin{matrix}
   \frac{\varepsilon_{0}k_{B}}{nq^{2}(\Delta x)^{2}}\left(T-T_{C}\right) & T>T_{C}  \\
   2\frac{\varepsilon_{0}k_{B}}{nq^{2}(\Delta x)^{2}}\left(T_{C}-T\right) & T<T_{C}  \\
\end{matrix} \right.
\end{equation}
\begin{equation}\label{EquPSM6}
  \left(2\pi\tau\right)^{-1}=\left\{ \begin{matrix}
   \left[\frac{\nu_{0}}{\pi}\left(\frac{T-T_{C}}{T}\right)\exp\left(-\frac{E_{A}}{k_{B}T}\right)\right] & T>T_{C}  \\
   2\left[\frac{\nu_{0}}{\pi}\left(\frac{T_{C}-T}{T}\right)\exp\left(-\frac{E_{A}}{k_{B}T}\right)\right] & T<T_{C}  \\
\end{matrix} \right.
\end{equation}
Here $\nu_{0}$ is the attempt frequency and $E_{A}$ is the energy barrier separating two local minima (Fig.~\ref{Phi4Model}).

Equations (\ref{EquPSM5}) and (\ref{EquPSM6}) correspond to an order
parameter which is homogeneous throughout the entire volume of the crystal. For a spatially
dependent order parameter, $P(\textbf{r})$, and assuming the Gaussian approximation, the relaxation time in Fourier-space becomes \cite{Nishimori_2011,Kardar_2007},
\begin{equation}\label{EquPSM7}
  \tau(q)^{-1}=\left(2a+bq^{2}\right)\Gamma \quad ,
\end{equation}
where $a=A(T-T_{C})$, $b$ is a constant and $q$ is the wave vector.  As a consequence, we see that each Fourier
component of the order parameter behaves as an independent particle connected to a spring
\cite{Kardar_2007} and the fluctuations in each mode decay with a different relaxation
time. Only in the long wavelength limit, $q=0$, the relaxation time in Eq.~(\ref{EquPSM7}) is equal to that in Eq.~(\ref{EquPSM6}).

In the opposite (displacive) limit of the model, $E_A/k_B T_C \ll 1$, the well-known expressions of Ginsburg-Landau mean-field theory are obtained for $q\rightarrow0$:
\begin{equation}\label{EquPSM5a}
  \Delta\varepsilon_{r}^{-1}=\left\{ \begin{matrix}
   \frac{2\varepsilon_{0}A}{C}\left(T-T_{C}\right) & T>T_{C}  \\
   \frac{4\varepsilon_{0}A}{C}\left(T_{C}-T\right) & T<T_{C}  \\
\end{matrix} \right.
\end{equation}
\begin{equation}\label{EquPSM6a}
  \left(2\pi\tau\right)^{-1}=\left\{ \begin{matrix}
   \frac{\Gamma_d A}{\pi} \left({T-T_{C}}\right) & T>T_{C}  \\
   \frac{2\Gamma_d A}{\pi} \left({T_{C}-T}\right) & T<T_{C}  \\
\end{matrix} \right.
\end{equation}
Here $A=3\theta k_B /(8 \pi J_0)$, $T_C=\eta/A$,  $\Gamma_d$ and $C$ are constants of the displacive limit;  $\theta$, and $\eta$ are microscopic parameters characterizing the nature of the potential in the $\phi^{4}$ single ion model
\cite{Bruce_1981,Giddy_1989JPCM,Salje_1991ACSA,Padlewski_1992JPCM,Radescu_1995JPCM}. Nota bene, in canonical ferroelectrics the ferroelectric phase transition of the displacive type is accompanied by softening of a characteristic phonon. In the long wavelength limit an overdamped softening mode is characterized according to Eq.~(\ref{EquPSM5a}) and Eq.~(\ref{EquPSM6a})\cite{blinc_book,Fujimoto_2010}.
Recently, a critical slowing down at the ferroelectric phase transition has been observed in a chiral multiferroic MnWO$_{4}$ [\onlinecite{Niermann2015}]. This behaviour has been attributed to a overdamped softening of an electromagnon mode\cite{pimenov_nphys_2006,Shuvaev2010PRB} obeying the temperature characteristic given in Eq.~(\ref{EquPSM5a}) and Eq.~(\ref{EquPSM6a}) with a critical exponent larger than 1.

\section{Results and Discussion}

\begin{figure}[tbp]
\begin{center}
\includegraphics[width=0.5\linewidth, clip]{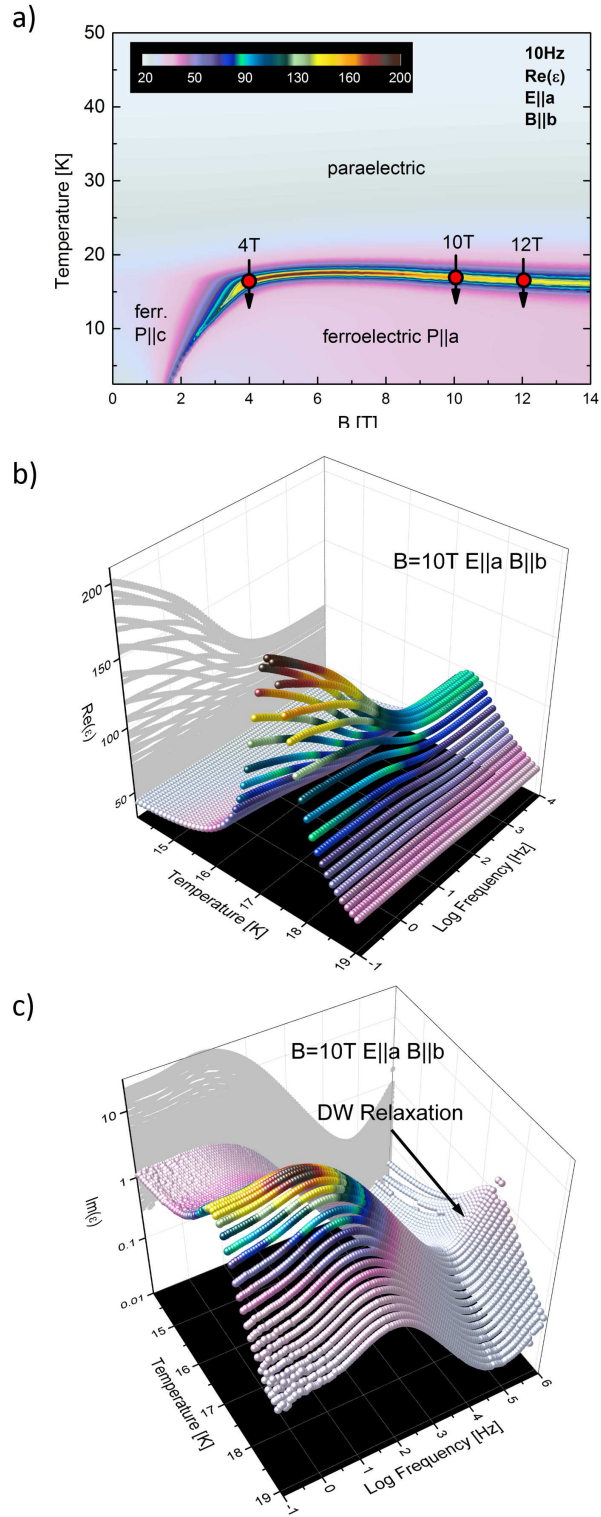}
\end{center}
\caption{\emph{Phase diagram of magnetoelectric \Dy~ in magnetic fields along the $b$-axis.} (a) The color map is a contour plot of the real part of the dielectric constant along the $a$-axis measured at 10 Hz. The ground state of \Dy~ in zero magnetic field is a $bc$-cycloid with $P\|c$. In external magnetic fields the electric polarization flops from $P\|c$ to $P\|a$. (b) Low frequency dispersion phenomena near $T_{C}$. (c) Logarithmic plot of the imaginary part of the permittivity reveals a well pronounced absorption in the vicinity of the paraelectric to ferroelectric phase transition and the signature of domain wall relaxation phenomenon at higher frequencys \cite{kagawa_prb_2011}.    } \label{PD}
\end{figure}

As demonstrated in Fig.~\ref{PD}a, the dielectric permittivity closely follows the known phase diagram of \Dy~ for $B \| b$-axis\cite{kimura_prb_2005}.  The changes in the dielectric permittivity are especially strong at the transition to the ferroelectric state with $P \| a$ (Fig.~\ref{PD}b). In addition a well pronounced absorption is observed in the close vicinity of $T_{C}$ (Fig.~\ref{PD}c)
Detailed analysis of the low-frequency dielectric relaxation is presented for magnetic fields
of 4T,10T and 12T corresponding to the transition from the sinusoidal paraelectric to the
$P\|a$-axis ferroelectric phase.

\begin{figure}[tbp]
\begin{center}
\includegraphics[width=0.5\linewidth, clip]{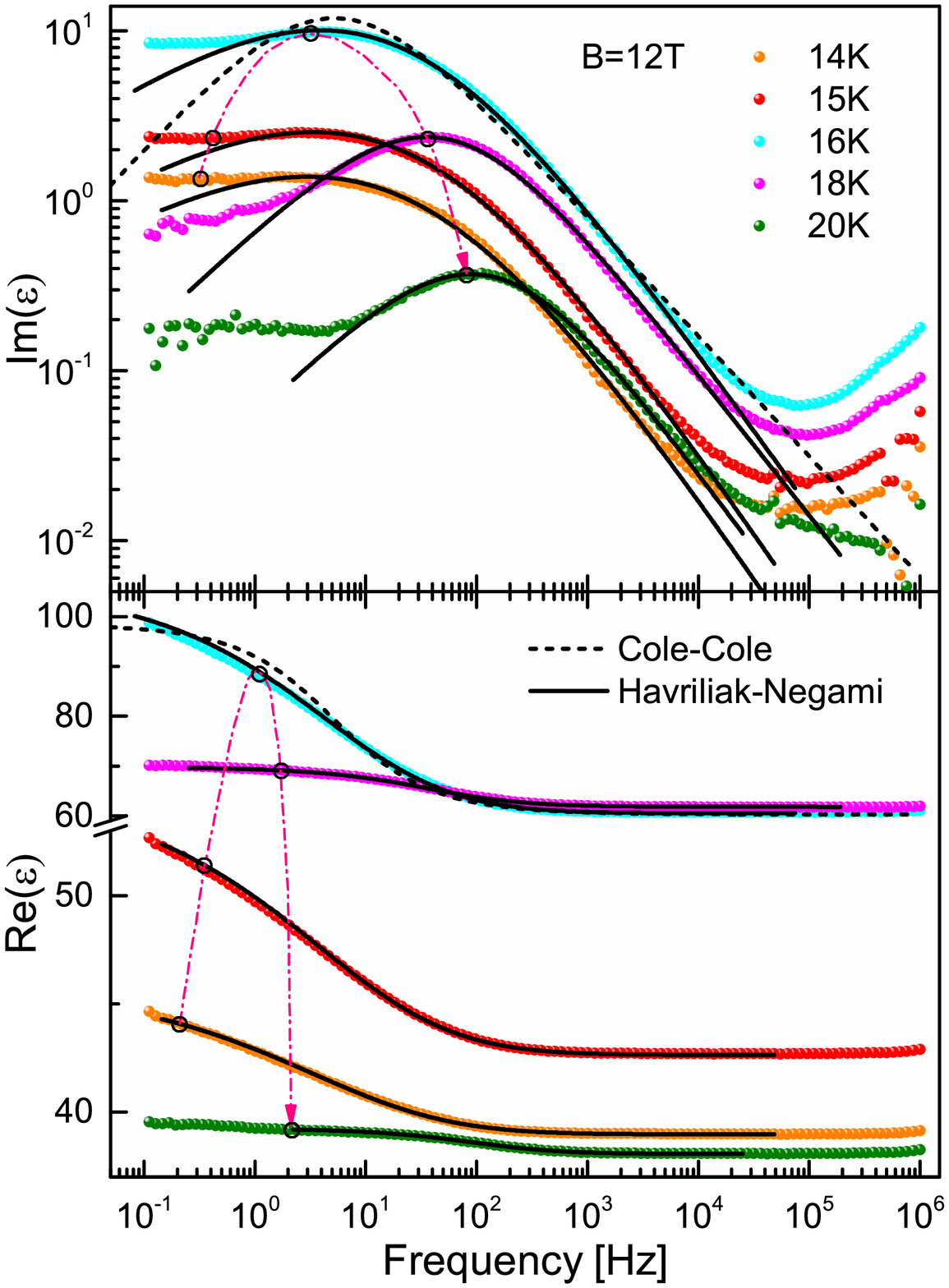}
\end{center}
\caption{\emph{Spectrum of the dielectric permittivity in a magnetic field of $\mu_{0}H=12$T.}
Typical frequency dependence of the dielectric constant, Re($\varepsilon$), and dielectric
loss, Im($\varepsilon$), at several temperatures above ($T=18$K, $T=20K$) and
below ($T=14K$, $T=15K$, $T=16K$) the critical temperature, $T_{C} \approx 17$~K (at 12~T). Black solids
lines represents the fit according to the Havriliak-Negami relaxation function, Eq.~(\ref{eqHN}). The dashed
black line shows the fit according to the Cole-Cole equation, Eq.~(\ref{eqCC}). Dash-dotted lines with open circles schematically indicate the position of the spectra with increasing temperature.} \label{fits}
\end{figure}

Figure~\ref{fits} shows typical dielectric spectra of \Dy~ close to the ferroelectric transition temperature $T_C \approx 18$~K. The spectra below the megahertz range are dominated by two relaxation processes. Only a wing of the high-frequency relaxation is seen in our spectra because the characteristic frequency of this mode is far above 1~MHz.  According to previous dielectric studies, the high-frequency mode can be attributed to the relaxation
of the domain walls~\cite{kagawa_prl_2009,kagawa_prb_2011}
in \Dy.

In the following we concentrate the analysis on an absorption peak observed for frequencies below 1 kHz.
As seen already in the spectra in Fig.~(\ref{fits},\ref{PD}b,\ref{PD}c), this peak grows in magnitude with decreasing temperature,
reaches a maximum value at $T_C$, and decreases again after passing the critical temperature.
The observed dielectric relaxation is slightly assymmetrical with broadening towards low-frequencies.

To obtain quantitative information of the origin of the low-frequency mode the spectra were fitted to the phenomenological Havriliak-Negami equation\cite{kremer_2002}:
\begin{equation}\label{eqHN}
  \varepsilon^{\prime}+\varepsilon^{\prime\prime}=
  \epsilon_{\infty}+\frac{\Delta\varepsilon}{(1+(i\omega\tau)^{1-\alpha})^{\beta}}
\end{equation}
Here $\Delta\varepsilon$ is the  relaxation strength, $\varepsilon_{\infty}$ is the  high frequency limit of the dielectric constant, $\tau$ is the characteristic relaxation time, $\alpha$ and $\beta$ are the width and asymmetry parameters, respectively. A simple Debye behaviour in Eq.~(\ref{eqHN}) would corresponds to $\alpha=0$ and $\beta=1$. Values of $0<\alpha<1$ and $0<\beta<(1-\alpha)^{-1}$ results in broadened asymmetric loss peaks with power laws of
$\omega^{1-\alpha}$ and $\omega^{-(1-\alpha)\beta}$ as the low- and high-frequency asymptotic behaviour, respectively.

Since the symmetric Cole-Cole function given by
\begin{equation}\label{eqCC}
  \varepsilon^{\prime}+\varepsilon^{\prime\prime}=
  \epsilon_{\infty}+\frac{\Delta\varepsilon}{(1+(i\omega\tau)^{1-\alpha})} \ ,
\end{equation}
is intensively used as a fitting function to describe permittivity data of conventional materials as well as magnetoelectric materials \cite{schrettle_prl_2009,Niermann2015}, we demonstrated by the dashed line in Fig.~\ref{fits} for the $T=16$~K data, that the symmetric Cole-Cole function results in a worse fit to the data compared to the Havriliak-Negami expression. Therefore, the subsequent analysis within the present work has been done according to  Eq.~(\ref{eqHN}). Most probably, a fitting procedure using the Cole-Cole function would lead to qualitatively similar behaviour of the relaxation time and dielectric strength.

The deviations from a Debye spectral shape of the relaxation are commonly ascribed to a distribution of relaxation times. According to Eq.~(\ref{EquPSM7}), this can be caused by fluctuations of the order parameter near the phase transition temperature. The mean
logarithmic relaxation time is related to the characteristic relaxation time by
\cite{Zorn_2002}
\begin{equation}\label{EquRD2}
  \langle\ln\tau_{HN}\rangle=\ln\tau+\frac{\psi(\beta)+Eu}{1-\alpha} \ ,
\end{equation}
where $\psi(\beta)$ is the digamma  function and $Eu \approx 0.577$ is the Euler constant. The width of a non-Debye relaxation is defined as the variance,
$\sigma^{2}=\langle(\ln\tau_{HN})^{2}\rangle-\langle\ln\tau_{HN}\rangle^{2}$, of the
distribution of logarithmic relaxation times and for a Havriliak-Negami function is given by\cite{Zorn_2002}
\begin{equation}\label{EquRD3}
  \sigma^{2}=\frac{\psi^{'}(\beta)}{(1-\alpha)^{2}}+\frac{\pi^{2}}{6(1-\alpha)^{2}}-\frac{\pi^{2}}{3} \ .
\end{equation}

\begin{figure}[tpb]
\begin{center}
\includegraphics[width=0.5\linewidth, clip]{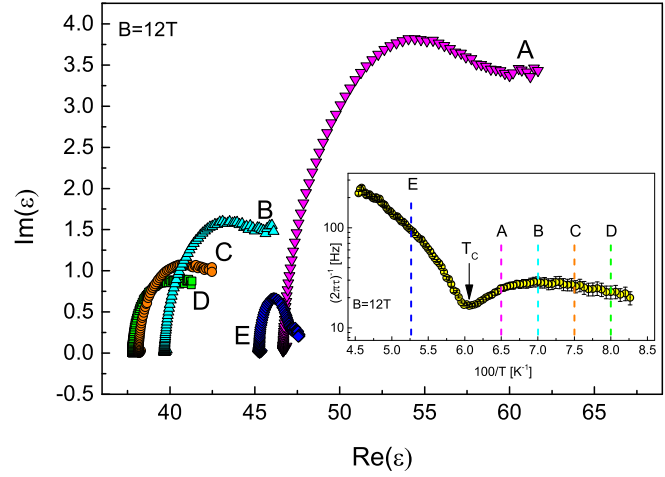}
\end{center}
\caption{\emph{Cole-Cole plots of low frequency relaxation.} The inset shows the temperature dependence of the relaxation time in the Arrhenius representation (see section \ref{sectime}).} \label{CCDW}
\end{figure}

In present work we do not consider the relaxations by the domain walls. Compared to fluctuations on the atomic level, ferroelectric domains are typically~\cite{fiebig_jpd_2005} of $\mu$m size and they are responsible for the high-frequency dielectric relaxation~\cite{kagawa_prl_2009,kagawa_prb_2011}. Further two arguments are in favor of nanosize origin of the relaxation discussed here: (i) the low frequency relaxation (Fig.~\ref{fits}) is well pronounced below and above $T_{C}$,  and (ii) no signature of thermally activated creep motion of domain walls is evident in the Cole-Cole plots (Fig.~\ref{CCDW}). In the latter case a linear relationship between the imaginary and real part of the permittivity with Im($\varepsilon$)$\propto \tan(\pi\beta/2)\cdot$Re($\varepsilon$) and $0<\beta <1$ is expected \cite{Braun_2005PRL,Kleemann_2002PRB,Kleemann_Ferooelectrics2006}.

\subsection{Relaxation Strength}
\begin{figure}[tbp]
\begin{center}
\includegraphics[width=0.5\linewidth, clip]{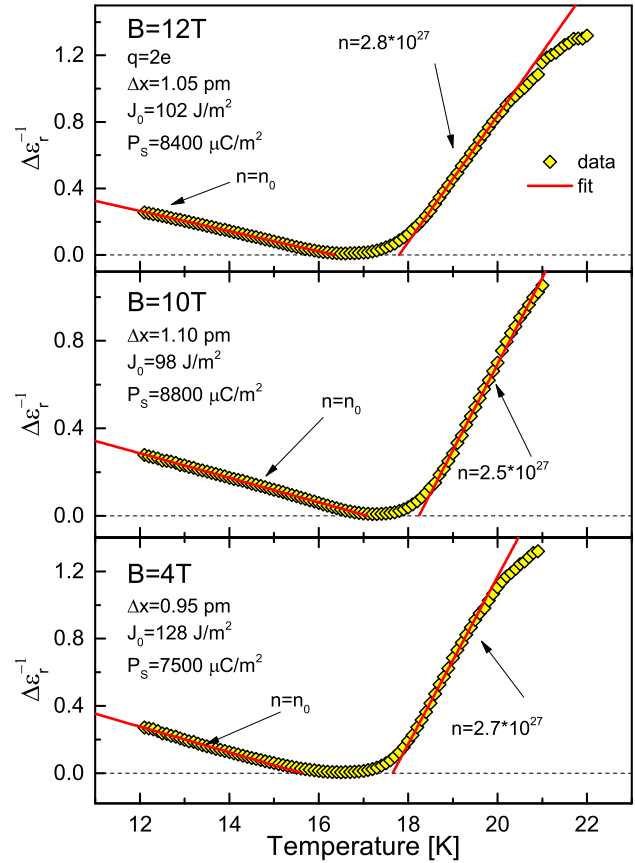}
\end{center}
\caption{\emph{Relaxation strength in \Dy.} Inverse relaxation strength along the $a$-axis of the low frequency mode as a function of temperature for different magnetic fields along the $b$-axis. The yellow symbols are obtained by the spectral analysis procedure and the red solid line correspond to fit function according to Eq.~(\ref{EquPSM5}).} \label{DeltaEps}
\end{figure}
The temperature dependence of  the inverse relaxation strength of the
low frequency mode is presented in Fig.~\ref{DeltaEps}.
Dielectric permittivity diverges as the temperature approaches $T_{C}$. At the ferroelecric transition temperature \Dy~ undergoes a
phase transition accompanied by a minimum in $\Delta\varepsilon_r^{-1}$ at $T_{C}$. As predicted by Eq.~(\ref{EquPSM5}), close to transition $\Delta\varepsilon_{r}^{-1}(T)$ shows the regions of linear dependence. This behaviour demonstrates critical dynamics characteristic for order-disorder phase transitions.

A significant rounding near $T_{C}$
is attributed to fluctuations of the order parameter and to limitation of the correlation length close to $T_{C}$ [\onlinecite{Lines}] . In addition, the straight lines of the model fits cross the x-axis at temperatures deviating by about one Kelvin.  These effects are not accounted within the present simple model since it implies a molecular field approach and neglects fluctuations.
Alternatively, a narrow temperature range near  $T_{C}$ may be analyzed using the formalism of critical exponents $\Delta \varepsilon^{-1} \propto |T-T_{C}|^{\gamma}$. Such analysis (not shown) reveals critical exponents for the relaxation strength of the low frequency mode near $T_{C}$ between
$\gamma=1.8$ and $\gamma=2.3$ depending on the magnetic field. These rather high values of the critical exponents may also be seen as an evidence for a order-disorder type phase transitions \cite{luban_jpc_1970}.


\begin{figure}[tbp]
\begin{center}
\includegraphics[width=0.5\linewidth, clip]{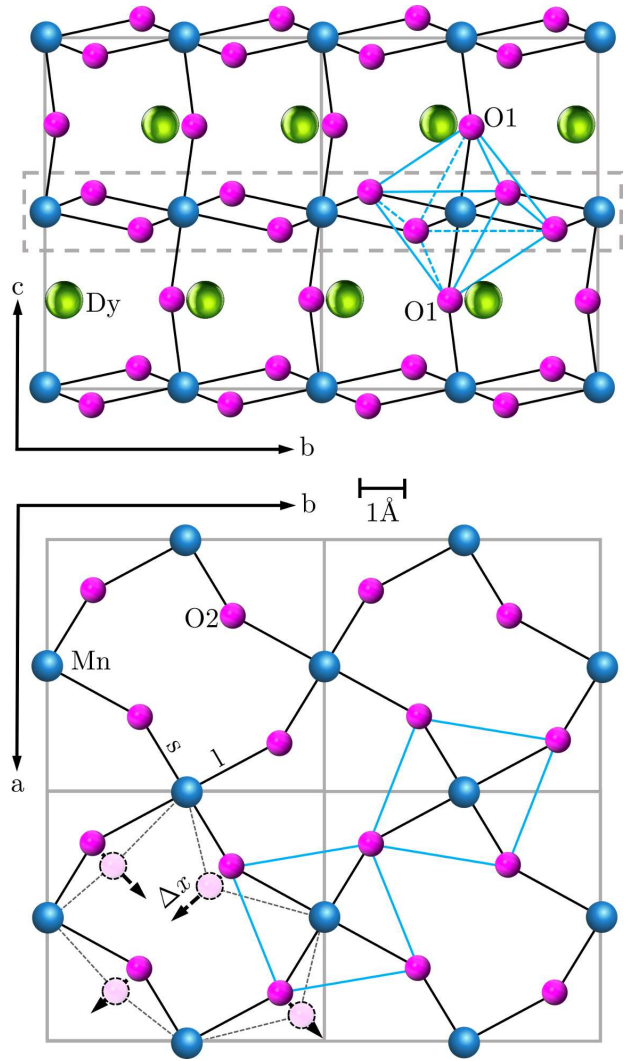}
\end{center}
\caption{\emph{Crystal structure of the orthorhombic rare earth manganite DyMnO$_{3}$.} Green spheres are Dy ions, blue spheres represent Mn ions and magenta spheres are O ions in Pbnm setting. Solid gray lines mark the unit cell. Lower frame shows the ab-plane of DyMnO$_{3}$ cut at the MnO$_2$ plane (dashed gray lines).} \label{Crystal}
\end{figure}

The electric dipole moment is generated by the displacements of the O$^{2-}$ ions and  only the ions labeled as O2 in Fig.~\ref{Crystal} generate a static electric polarization along the $a$-axis. Therefore, we assume an electric dipole density in \Dy~ equal to $2/3$ of the density of O$^{2-}$ ions in the unit, $n_{O^{2-}}=n_{0}\sim 3.5\cdot 10^{28}m^{-3}$, [\onlinecite{muthu_jpconf_2012,Alonso_1999}] and a charge of $q=2e$, where $e$ is the elementary charge.
According to Eq.~(\ref{EquPSM5}), the slope of $\Delta\varepsilon_{r}^{-1}$ is inversely proportional to $n(q\Delta x)^2=2P_s^2/n$ with static electric polarization along the a-axis $P_s=nq\Delta x / \sqrt{2}$. The factor $\sqrt{2}$ appears to to 45$^{\circ}$ degree misalinement between the oxygen displacement and the a-axis (Fig.~\ref{Crystal}). Hence, for $T<T_C$ from the slopes of  $\Delta\varepsilon_{r}^{-1}$ we may directly estimate the static electric polarization. The microscopic parameters of the model, $\Delta x$ and $J_0$ are obtained from the values of the static polarization and from $J_{0}=k_{B}T_{C}/2(\Delta x)^{2}$. These parameters are given in Fig.~\ref{DeltaEps}. Taking into account the simplicity of the model, the obtained values of the electric polarization agree reasonably well with directly measured data~\cite{kimura_prb_2005} $P\sim 2000 \mu C/m^2 $. In addition, the obtained coupling constant coincide with the classical electric dipole energy in the Mn-O-Mn chains along the b-axis $J_{0}\sim (q^{2}/2\pi \varepsilon_{0})/(l^{2}+s^{2})^{3/2}\sim70 J/m^2$ (Fig.~\ref{Crystal}) with $l=2.22\AA$ and $s=1.90\AA$ being the long and short bond distances~\cite{Alonso_1999} .

The temperature dependence of the inverse relaxation strength becomes steeper in the paraelectric phase $T>T_C$ although the model predicts an opposite behavior. This effect is evidently not captured within the assumptions of present simple model. Two possible explanations for this behaviour may be suggested in present stage: (i) Change in effective dipole density at the phase transition. In the disordered state a large portion of the magnetic cycloid is distorted and is included into the border regions between the left- and right-rotating cycloids. The oxygen ions in these regions are effectively excluded from the relaxation process thus reducing the effective dipole density $n$ of the model. This explanation has been used in the fits to the data in Fig.~\ref{DeltaEps} and the effective dipole density for $T > T_C$ is indicated at the fit lines. (ii) The model can be modified to account for higher order terms \cite{CanoPr,CanoPRB_2010}. A coupling term between the polarization and magnetization may me explicitly included into the free energy expansion, Eq.~(\ref{EquPSM4}). The term of which is always allowed by symmetry is $\gamma P^{2}M^{2}$. Thus the free energy expansion near $T_{C}$, Eq.~(\ref{EquPSM3}), can be written as
\begin{equation}\label{FME}
 F=aP^{2}+bP^{4}-gEP+\gamma P^{2}M^{2} \ ,
\end{equation}
where $a=A(T-T_{C})$, and $b, g$, and $\gamma$ are the constants of the model. In these modification, $M$ has to be understood as the amplitude of the transverse component of the spin-cycloidal $S=(0,M\cos qy, M\sin qy)$.
Applying Eq.~(\ref{FME}) to Eq.~(\ref{EquPSM1}) leads to
\begin{equation}\label{FME2}
\frac{\partial P\left(t\right)}{\partial t}=-\Gamma\left(2aP+4bP^{3}-gE+\gamma PM^{2}\right) \ .
\end{equation}

With $E=\delta E\exp\left(-i\omega t\right)$ and $P=P_{S}+\delta P\exp\left(-i\omega t\right)$ and assuming that $M^{2}\propto\left(T_{C}-T\right)=-Ca$ below $T_{C}$ and $M^{2}=0$ above $T_{C}$ the relaxation strength and relaxation time becomes
\begin{equation}\label{FME3}
  \Delta\varepsilon_{r}^{-1}=\left\{ \begin{matrix}
   \frac{\varepsilon_{0}k_{B}}{nq^{2}(\Delta x)^{2}}\left(T-T_{C}\right) & T>T_{C}  \\
   2(1-\gamma C)\frac{\varepsilon_{0}k_{B}}{nq^{2}(\Delta x)^{2}}\left(T_{C}-T\right) & T<T_{C}  \\
\end{matrix} \right.
\end{equation}
\begin{equation}\label{FME4}
  \left(2\pi\tau\right)^{-1}=\left\{ \begin{matrix}
   \left[\frac{\nu_{0}}{\pi}\left(\frac{T-T_{C}}{T}\right)\exp\left(-\frac{E_{A}}{k_{B}T}\right)\right] & T>T_{C}  \\
   2(1-\gamma C)\left[\frac{\nu_{0}}{\pi}\left(\frac{T_{C}-T}{T}\right)\exp\left(-\frac{E_{A}}{k_{B}T}\right)\right] & T<T_{C}  \\
\end{matrix} \right.
\end{equation}

It is clear, that adding explicitly magnetoelectric terms into the free energy, the ratio of the slopes of the relaxation strength below and above $T_{C}$ deviate from the value of 2 as predicted by a simple model. In this case, the slope below $T_{C}$ is governed by a new parameter $(1-\gamma C)$ in Eqs.~(\ref{FME3},\ref{FME4}).

\subsection{Relaxation Time} \label{sectime}

\begin{figure}[tbp]
\begin{center}
\includegraphics[width=0.5\linewidth, clip]{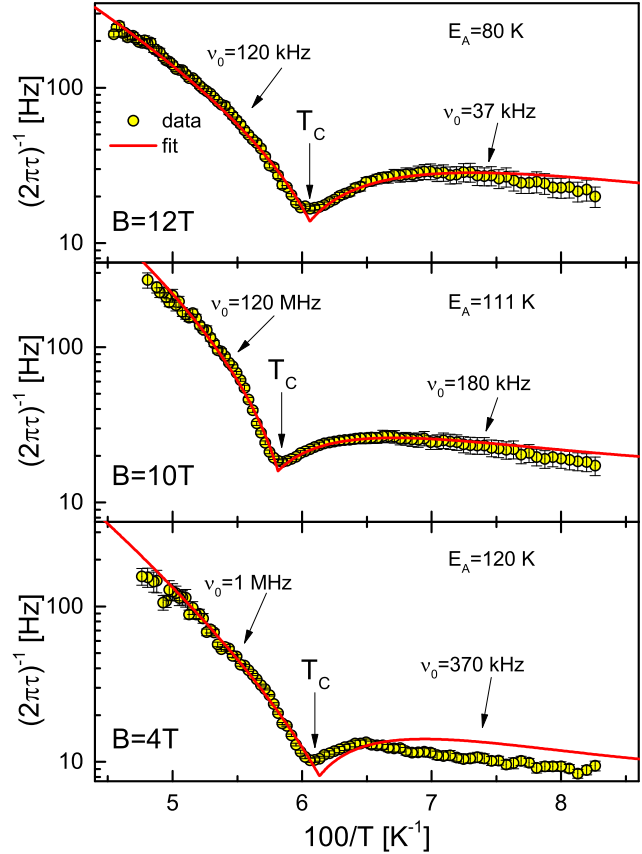}
\end{center}
\caption{\emph{Relaxation time in \Dy.}  Inverse mean relaxation time along the $a$-axis of the low frequency mode as a function of temperature for different magnetic fields $B \| b$-axis.
The yellow symbols are obtained by the spectral analysis procedure and the red solid line correspond to fit function according to Eq.~(\ref{EquPSM6}).} \label{RelaxTime}
\end{figure}

The inverse relaxation time of the low frequency mode obtained
from the spectral analysis is
shown in Fig.~\ref{RelaxTime}. On cooling from the paraelectric phase the inverse relaxation time decreases toward $T_{C}$. Below $T_{C}$, $1/ \tau$ shows a broad characteristic maximum and decreases again for low temperatures. Qualitatively, the temperature behavior of the relaxation time can be explained as a superposition of two processes: (i) Activated behaviour with a characteristic energy of $E_a \sim 100$~K and (ii) Critical slowing down of the relaxation in the vicinity of $T_C$. This observation is typical for order-disorder phase transitions involving
shallow double well potentials \cite{Stareinic2006_241,Lines,Gonzalo}.

Both processes determining the temperature evolution of the relaxation time are qualitatively well captured within the present simple model.  The temperature activated
behaviour, expressed by the exponential factor in Eq.~(\ref{EquPSM6}), prevails for
temperatures far from $ T_{C}$ and, therefore, causes an overall decrease of the relaxation time for decreasing temperature.
Qualitatively similar behaviour of the relaxation time was found\cite{schrettle_prl_2009} for a c-axis relaxation ($e \| c$) in \Dy~ for a transition to a $bc$-cycloidal magnetic ordering.
The $P \| c$-axis state is achieved in Dy for cooling in zero external magnetic field. It seem to be plausible that the c-axis relaxation in multiferroic manganites may be explained by the present model as well.

In case of inverse relaxation time the suggested model gives a qualitative explanation of the observed data. Eq.~(\ref{EquPSM6}) contains two temperature-dependent factors, Arrhenius term $exp(-E_A/(k_B T))$ and the critical-slowing term $\nu_0 (T-T_C)/T$. These two terms qualitatively explain the temperature dependence of the relaxation time close to phase transition.  In order to obtain reasonable fits to the critical behavior of the relaxation time, different values of the attempt frequency above and below $T_C$ have been used as well as a constant was added to Eq.~(\ref{EquPSM6}). Nota bene, accounting magnetodielectric effects the slope of the inverse relaxation time is not necessarily 2, see Eq.~(\ref{FME4}). Thus the values of the attempt frequency above and below $T_C$ can take the same value. Eventually, in addition to magnetodielectric effects, the distribution of relaxation times plays also an role to explain that feature.

The ratio of $E_{A}/k_B T_{C}\sim 5$ is a further evidence of an order-disorder type phase transition\cite{Dove_1997} and is in contrast to a displacive type phase transition where $E_{A}/k_{B}T_{C}\ll1$ holds. Furthermore,  $E_A / k_B \sim 100$~K  corresponds well to a characteristic energy of the magnetic order as determined by N\'{e}el temperature of $T_N \sim39$K [\onlinecite{Aubry_JCP1975}].

\begin{figure}[tbp]
\begin{center}
\includegraphics[width=0.5\linewidth, clip]{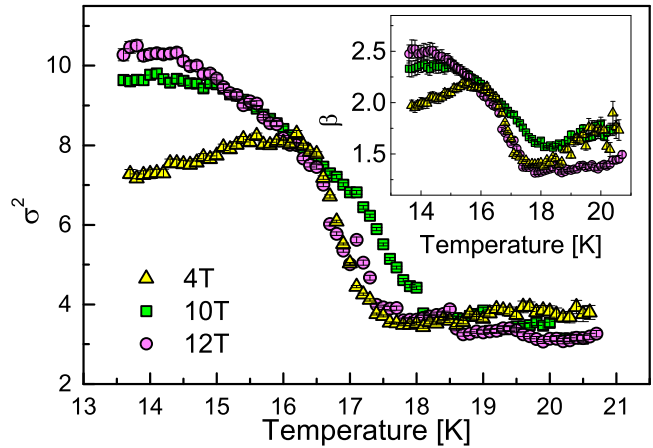}
\end{center}
\caption{\emph{Relaxation width in \Dy.}  Asymmetry and width of the low-frequency dielectric relaxation in \Dy, obtained via Eq.~(\ref{EquRD3})} \label{ShapePara}
\end{figure}

Finally, Fig.~\ref{ShapePara} shows the width of the low-frequency dielectric relaxation in \Dy, obtained via Eq.~(\ref{EquRD3}). The increase of the characteristic width below $T_C \sim 18$~K is clearly seen. In the ordered magnetic state the effective length of the elementary cycloids increases thus leading to broader length distribution. Most probably, this also leads to the observed broadening of the dielectric relaxation.

\section{Discussion}

The analysis of the low-frequency dielectric relaxation demonstrates an overlap of two processes: critical dynamics close to $T_C$ and activation behavior in the broader frequency range. This observation may be reasonably explained using a simple model of an order-disorder phase transition with a double well potential. This potential reflects a dynamical switching between cycloids of the opposite chirality. Several parameters of the model correlate well with physical properties of \Dy. Thus, the characteristic energies of magnetic ordering and the value of the static electric polarization are in agreement with known values. We note that present experiments could be analyzed up to $T = 22$~K only. Strictly speaking, the dynamical behaviour of the cycloids still may change close to $T_N$.

Most importantly, the present experimental data and the simple model suggest to explain the paraelectric sinusoidal phase in rare-earth manganate as a dynamical equilibrium of cycloids with opposite chiralities. In addition to the dielectric results, this hypothesis resolve several experimental constraints which contradicted the concept of static sinusoidally modulated magnetic phase.

\section{Conclusions}

Low-frequency relaxation mode is observed in dielectric properties of \Dy~ multiferroic manganite and it reveals critical behaviour at ferroelectric transition temperature, $T_C \sim 18$~K. Together with temperature-activated relaxation rate, the observed mode may be qualitatively explained within a model for order-disorder phase transition. The model assume a switching between magnetic  cycloids with opposite chirality and correlates well with known physical properties of \Dy.

Combining present results with several other experiments on multiferroics we suggest that the paramagnetic sinusoidal phase should be explained as a dynamical equilibrium between the clockwise and counterclockwise cycloidal magnetic orders. The short range order in the paraelectric phase is transformed to a long-range cycloid at the ferroelectric transition temperature.

\subsection*{Acknowledgements}

This work was supported by by the Austrian Science Funds
(I815-N16, W1243).

\bibliography{literature}

\end{document}